\documentclass{PoS}

\title{Lattice Studies for hadron spectroscopy and interactions}

\ShortTitle{Lattice Studies for hadron spectroscopy and interactions}

\author{\speaker{Sinya Aoki}
\\
        Yukawa Institute for Theoretical Physics, Kyoto University, Kitashirakawa Oiwakechou, Sakyo-ku, Kyoto 606-8502, Japan \\
        E-mail: \email{saoki@yukawa.kyoto-u.ac.jp}}

\abstract{Recent progresses of lattice QCD studies for hadron spectroscopy and interactions are briefly reviewed. Some emphasis are given on a new proposal for a method, which enable us to calculate potentials between hadrons. As an example of the method, the extraction of nuclear potential in lattice QCD is discussed in detail.}

\FullConference{XV International Conference on Hadron Spectroscopy-Hadron 2013\\
		4-8 November 2013\\
		Nara, Japan }

\begin{document}

\section{Introduction: Current status}
\subsection{Hadron spectroscopy}
\begin{figure}[tbh]
\begin{center}
  \includegraphics[width=0.45\textwidth]{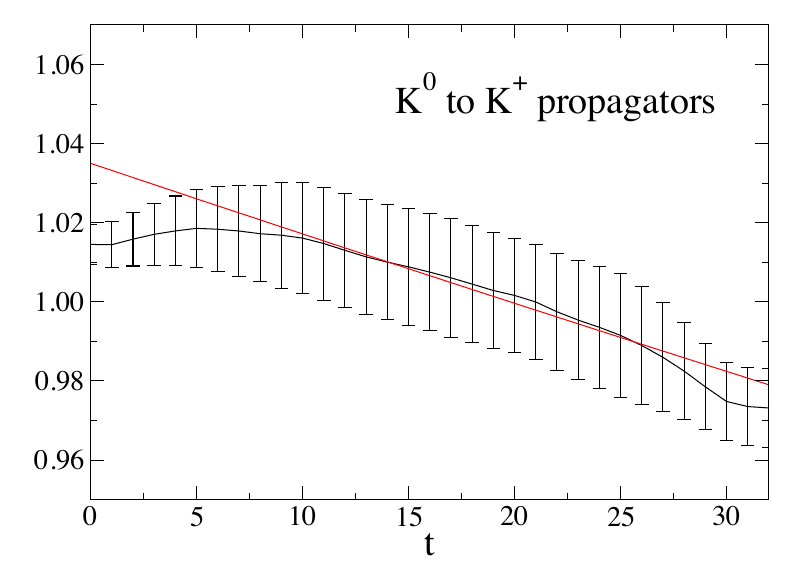}
 \vskip -0.2cm
\caption{A ratio of $K^0$ to $K^+$  propagators as a function of $t$.  The results (black symbol)
are consistent with the expected slope from the experimental value (red- line).  A figure is taken from Ref.~\cite{Aoki:2012st}. }
\vskip -0.6cm
\label{fig:K0K+}
\end{center}
\end{figure}

As an introduction of my presentation,  let me show one figure, which well represents a recent status of hadron mass calculations in  lattice QCD.  Fig.~\ref{fig:K0K+} shows a ratio of $K^0$ (neutral $K$ meson) to $K^+$ (charged $K$ meson) propagators as a function of $t$, which therefore behavior as $e^{-(m_{K^0} - m_{K^+})t} \simeq 1-(m_{K^0} - m_{K^+})t$ at large $t$ but small $(m_{K^0} - m_{K^+})t$. The fit to data gives $m_{K^0}-m_{K^+} = 4.54(1.09)$ MeV, which agrees with the experimental value, $m_{K^0}-m_{K^+} = 3.937(28)$ MeV,  within a large error.  This result has been obtained by 1+1+1 flavor QCD+QED simulations at the physical quark masses, where effects of both up-down quark mass difference and the dynamical QED  are included by the reweighting method\cite{Aoki:2012st}, though the continuum limit has not been taken yet.  As represented by this figure, isospin breaking effects due to both mass and charge differences of up-down quarks can be included in current lattice QCD calculations.   

\subsection{Hadron interactions}
In lattice QCD,  hadron interactions can be investigated  by  three methods so far,
two of which are standard ones and are briefly discussed here. 

The most straightforward method is to calculate nuclei directly in lattice QCD. 
This method is the most ab-initio but very difficult, since a number of contractions for quark fields increases naively as $(3A)!$ where $A$ is an atomic number of a target nucleus\footnote{A method to reduce this number drastically has been proposed recently\cite{Doi:2012xd}.}. Moreover  a signal to noise ratio  of the nucleus propagator decreases as $e^{-(E_A - 3A m_\pi) t}$ for large $t$,  where $E_A = O(A m_N)$ is the ground state energy of the nucleus and $m_\pi$ ($m_N$) is pion (nucleon) mass. 
Because of these difficulties, current studies are limited for light nuclei.
It is also difficult to apply a result obtained for one system to other systems.

\begin{figure}[tbh]
\begin{center}
  \includegraphics[width=0.9\textwidth]{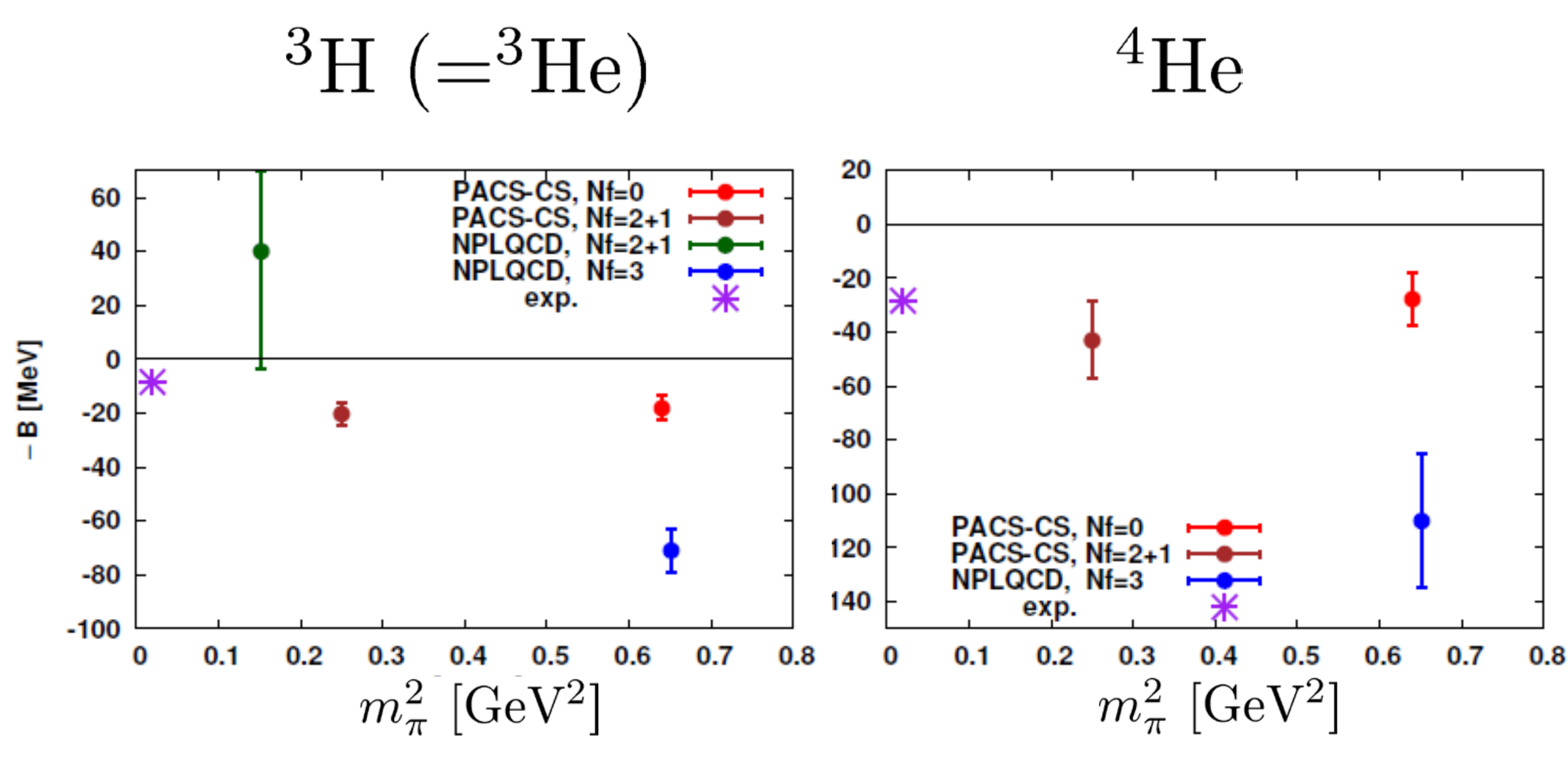}
  \vskip -0.3cm
\caption{Compilations of binding energy $\Delta E$ as a function of $m_\pi^2$ for $^3$H (left) and $^4$He (right).
 Figures are taken from Ref.~\cite{Doi:2012ab}. }
 \vskip -0.5cm
\label{fig:helium}
\end{center}
\end{figure}

Fig.~\ref{fig:helium} shows compilations of recent results for binding energies of $^3$H (triton) and $^4$He (helium 4) as a function of pion mass squared $m_\pi^2$, where 
solid triangle\cite{Yamazaki:2009ua} and circle\cite{Yamazaki:2012hi} represent results obtained by  PACS-CS collaboration for quenched and 2+1 flavor QCD, respectively,  
while open circle\cite{Beane:2010em} and square\cite{Beane:2012vq} represent
those obtained by NPL QCD collaboration for 2+1 and 3 flavor QCD.
Although results from two groups are largely scattered, an order of magnitude of binding energies turns out to be comparable to experimental values. Careful investigations  on systematic errors of course will be needed for more reliable estimation of these binding energies.      

The standard method to investigate hadron interactions in lattice QCD is to calculate scattering phase shift by the L\"usher's finite volume method\cite{Luscher:1990ux}, which relates spectra of two particles in the finite box to scattering phases of these two particles in the infinite volume. This method gives ab-initio calculations for phase shifts.  Once nucleon-nucleon ($NN$) scattering phase shifts have been calculated at a sufficiently wide range of energies by this method, for example, one can construct nuclear potentials form these results, which can be used to calculate nuclear structures  by solving many-body Schr\"odinger equations.

\begin{figure}[bth]
\begin{center}
\vskip -0.5cm
  \includegraphics[width=0.5\textwidth]{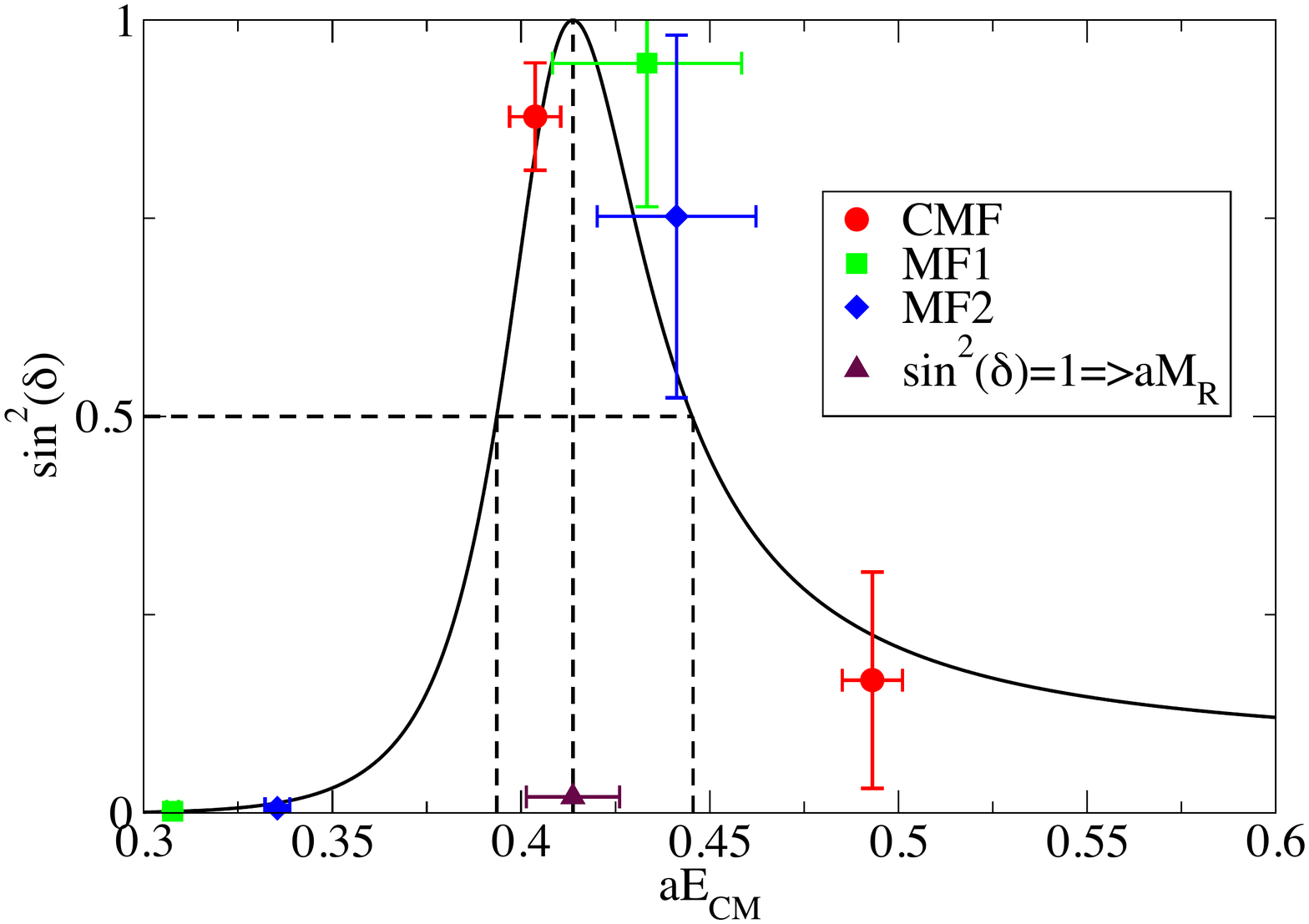}\hfill
  \includegraphics[width=0.5\textwidth]{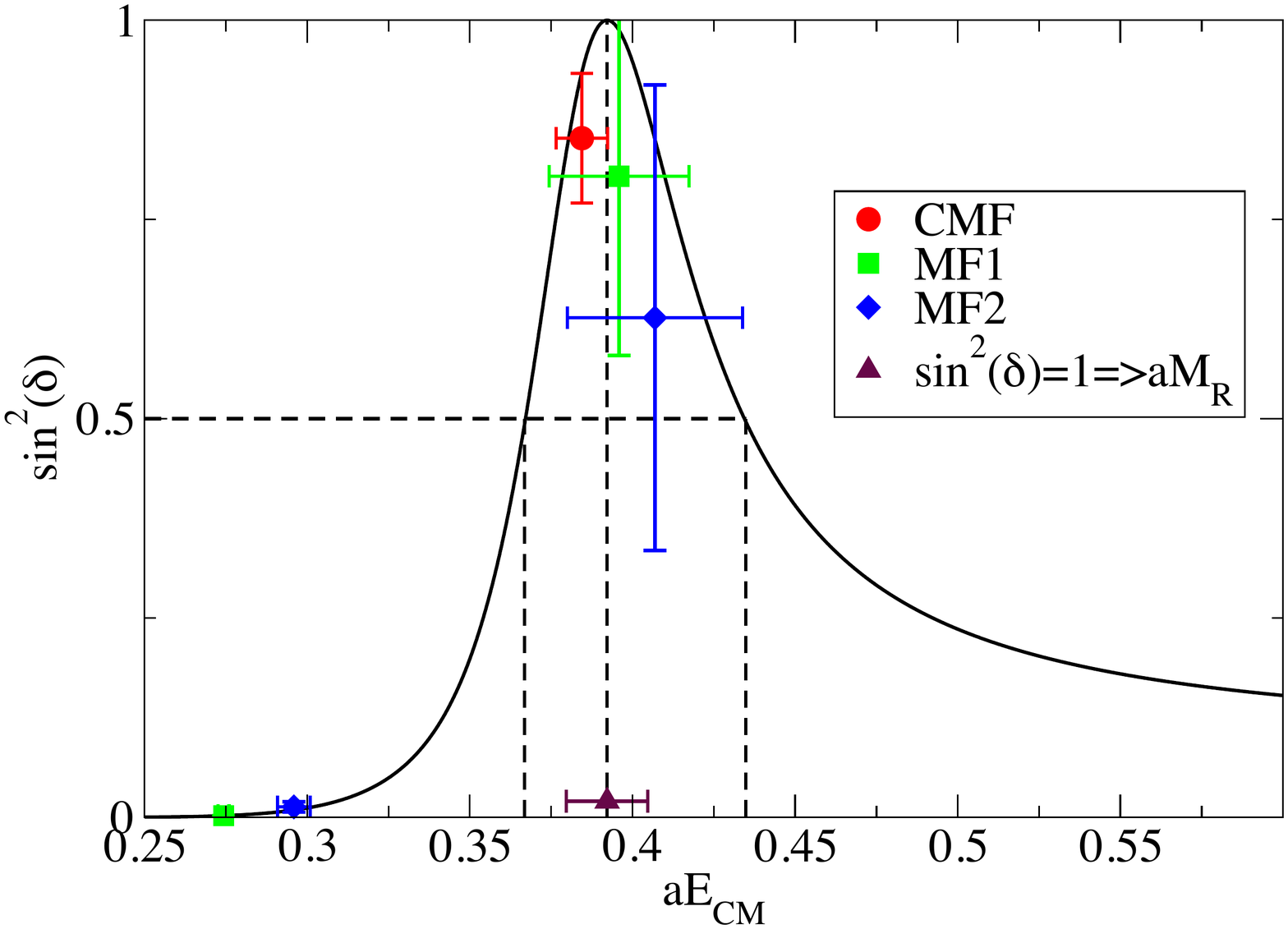}
  \vskip -0.5cm
\caption{Phase shift of the $\pi^+\pi^-$ scattering as a function of $a E_{\rm CM}$, together with the fit line to the effective range formula, at $m_\pi \simeq 330$ MeV (left) and 290 MeV (right). 
 Figures are taken from Ref.~\cite{Feng:2010es}. }
 \vskip -0.5cm
\label{fig:rho-pipi}
\end{center}
\end{figure}

Fig.~\ref{fig:rho-pipi} plots $\sin^2\delta$ as a function of the center of mass energy in lattice unit, $a E_{\rm CM}$ , where $\delta$ is  the phase shift of the $\pi^+\pi$-scattering in  the isospin $I=1$ and the orbital angular momentum $L=1$ channel, at $m_\pi \simeq 330$ MeV (left) and $m_\pi \simeq 290$ MeV (right). 
The solid line in the figure represents the fit to data by the effective range formula, 
\begin{equation}
\tan \delta = \frac{g_{\rho\pi\pi}^2}{6\pi}\frac{(E_{\rm CM}^2/4-m_\pi^2)^{3/2}}{E_{\rm CM}(m_\rho^2-E_{\rm CM}^2)}
\end{equation}
where $m_\rho$ is a "mass" of the $\rho$ meson resonance  and $g_{\rho\pi\pi}$ is the effective $\rho\rightarrow \pi\pi$ coupling constant. Although the pion is still heavier than the physical one, this results shows that hadron resonances can be treated in current lattice QCD simulations. 

\section{HAL QCD method to hadron interactions}
 An alternative method to the previous two standard methods has been proposed recently to investigate hadron interactions, and is called HAL QCD method\cite{Ishii:2006ec,Aoki:2009ji,Aoki:2011ep,Aoki:2012tk}, which gives the ab-initio extraction for a potential between two hadrons below inelastic threshold in QCD. The resultant potential has a clear physical interpretation and can be employed for many-body Schr\"odinger equations to investigate nuclear structures. We first explain the strategy of the HAL QCD method.
 
\subsection{Strategy}
 A basic but important quantity is the Nambu-Bethe-Salpeter (NBS) wave function\cite{Balog:2001wv}, which is defined, for example, for two nucleons in QCD as
\begin{eqnarray}
\Psi_{\vec k} (\vec x) &=& \langle 0 \vert T\{ N(\vec r,0) N(\vec r + \vec x, 0) \}\vert NN, W_k \rangle_{\rm in}
\end{eqnarray}
where $\langle 0 \vert$ is the QCD vacuum state, $\vert NN, W\rangle_{\rm in}$ is a two nucleon asymptotic in-state with the total energy $W_k=2\sqrt{\vec k^2 +m_N^2}$, the nucleon mass $m_N$ and a relative momentum $\vec k$  in the center of mass system, $T$ means the time ordered product, and $N(x)$ with $x= (\vec x, t)$ is a nucleon operator.
 
 As $x=\vert\vec x\vert$ becomes large, the NBW wave function satisfies the free Scgr\"odinger equation, 
 $(E_k - H_0) \Psi_{\vec k}(\vec x) \simeq 0$, where $E_k =\vec k^2/(2\mu)$, $H_0=-\nabla^2/(2\mu)$ with the reduced mass $\mu=m_N/2$. Furthermore, an asymptotic behavior of the NBS wave function can be determined in terms of the phase $\delta$ whose existence is implied by the unitarity of the $S$-matrix in QCD\cite{Lin:2001ek,Aoki:2005uf,Ishizuka2009a}: 
\begin{eqnarray}
\Psi_{\vec k}^L(x) &\simeq& A_L \frac{\sin(kx -L\pi/2+\delta_L(W_k))}{kx},\quad k=\vert\vec k\vert
\end{eqnarray}
at
$W_k < W_{\rm th} \equiv 2m_N + m_\pi$ for the partial wave with the orbital angular momentum $L$.

The HAL QCD method is based on  an existence of  a non-local but energy independent potential $U(\vec x, \vec y)$ which satisfies 
\begin{eqnarray}
(E_k - H_0) \Psi_{\vec k}(\vec x) &=& \int d^3y\, U(\vec x, \vec y) \Psi_{\vec k}(\vec y),
\end{eqnarray}
where the energy-independence means $U$ does not depend on the energy $W_{\vec k}$ of a particular NBS wave function. An existence of such $U$ can be shown by explicitly constructing it as
\begin{eqnarray}
U(\vec x,\vec y) &=& \sum_{\vec k,\vec k^\prime}^{W_k, W_{k^\prime} < W_{\rm th}}
(E_k - H_0) \Psi_{\vec k}(\vec x) \eta_{\vec k,\vec k^\prime}^{-1} \Psi_{\vec k^\prime}^{\dagger}(\vec y) ,
\end{eqnarray}
where $\eta_{\vec k,\vec k^\prime}^{-1} $ is an inverse of $\eta_{\vec k,\vec k^\prime} = \int d^3x\, \Psi_{\vec k}^\dagger (\vec x) \Psi_{\vec k^\prime}(\vec x) $ in the space spanned by $\{ \Psi_{\vec k}, W_k < W_{\rm th} \}$.
Indeed it is easy to see
\begin{eqnarray}
\int d^3y\, U(\vec x, \vec y) \Psi_{\vec p}(\vec y) &=&  \sum_{\vec k,\vec k^\prime}^{W_k, W_{k^\prime} < W_{\rm th}}
(E_k - H_0) \Psi_{\vec k}(\vec x) \eta_{\vec k,\vec k^\prime}^{-1}  \eta_{\vec k^\prime,\vec p} = (E_p -H_0)
\Psi_{\vec p}(\vec x) 
\end{eqnarray}
for $^\forall \vec p$ with $W_p < W_{\rm th}$. 

For practical uses, this non-local potential is expanded in terms of derivatives as $U(\vec x, \vec y) =V(\vec x,\vec \nabla) \delta^{(3)}(\vec x-\vec y)$, which is truncated at lowest few orders. For example, the leading order potential is simply given by
\begin{eqnarray}
V_{\rm LO}(\vec x) &=& \frac{(E_k - H_0)\Psi_{\vec k}(\vec x)}{\Psi_{\vec k} (\vec x)},
\label{eq:LO}
\end{eqnarray}
where $V_{\rm LO}(\vec x)$ depends on a particular choice of the NBS wave function, $\Psi_{\vec k}(\vec x)$,
due to the truncation of the derivative expansion at the leading order.
In the HAL QCD method, once the potential is obtained, physical observables such as scattering phase shifts and  energies of possible bound states can be extracted by solving Schr\"odinger equation with this potential.

\subsection{Nuclear potentials in lattice QCD}
\begin{figure}[tbh]
\begin{center}
\vskip -0.7cm
  \includegraphics[width=0.5\textwidth]{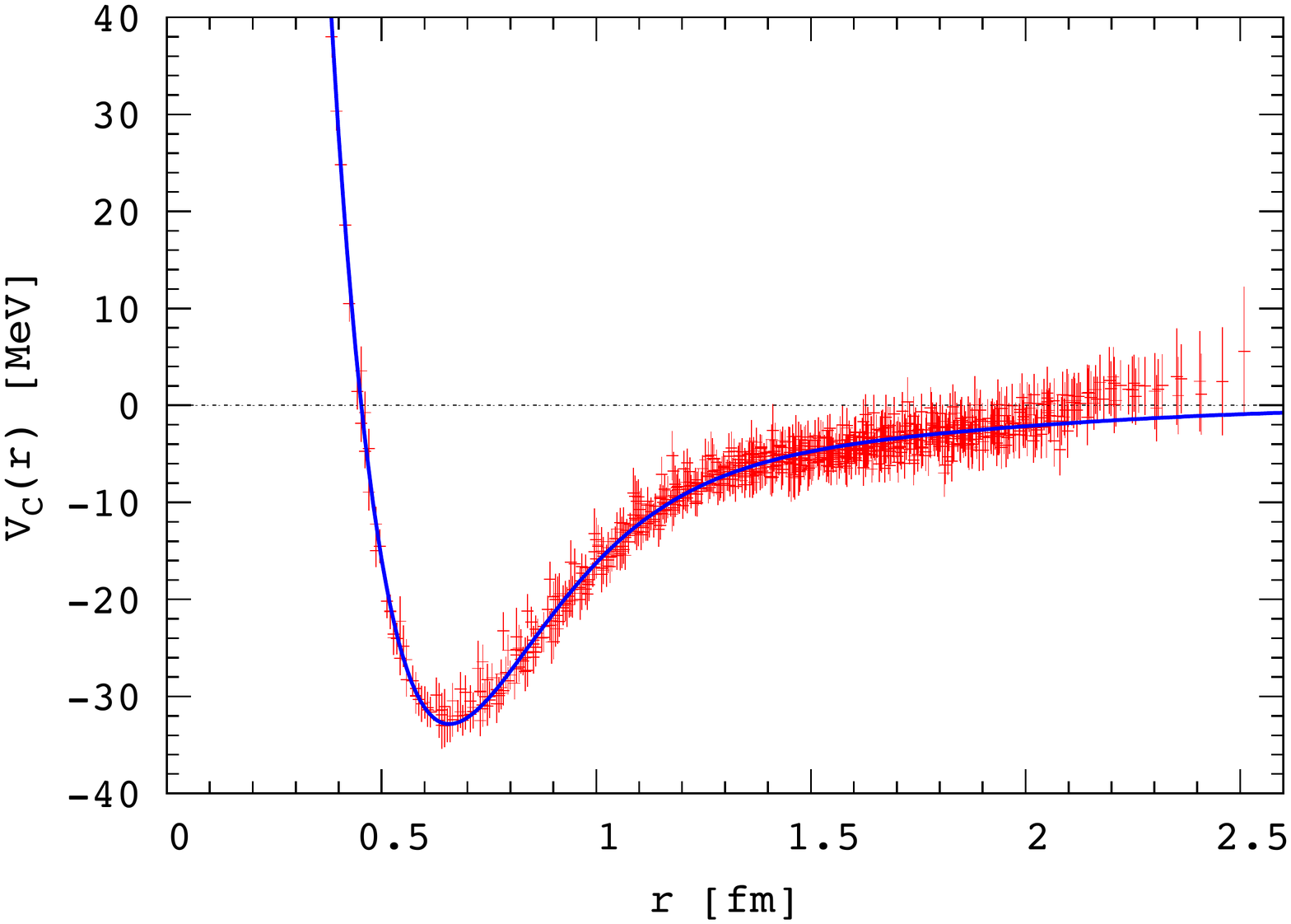}\hfill
  \includegraphics[width=0.5\textwidth]{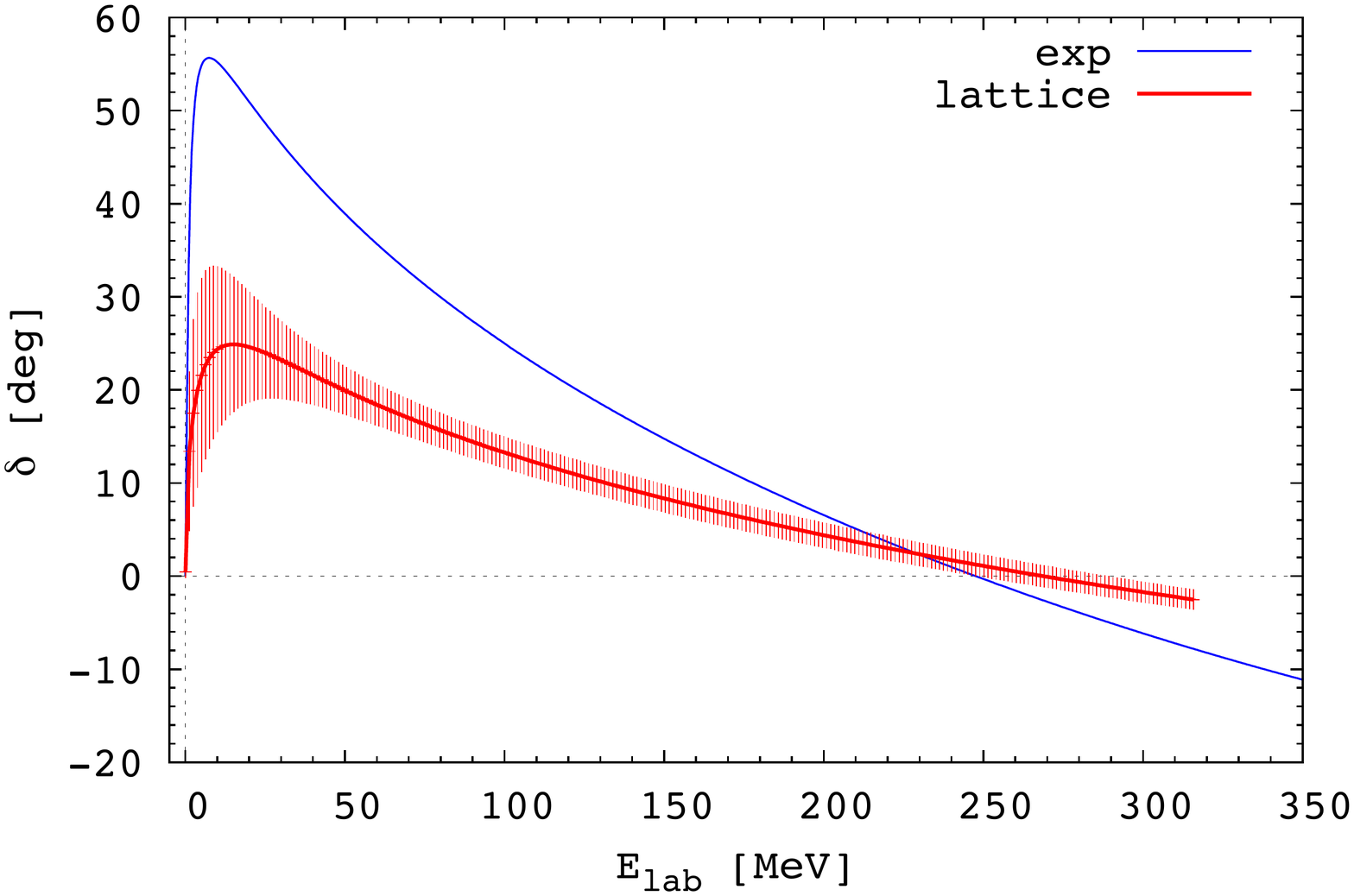}
  \vskip -0.5cm
\caption{(Left) The isospin-triplet NN central potential $V_C$ at the leading order of the derivative expansion as a function of $r=\vert\vec{x}\vert$ with the multi-Gaussian fit by the  solid line. (Right)  The scattering phase shift in $^1S_0$ channel as a function of the laboratory energy $E_{\rm lab}$, together with experimental data\cite{nn-online}. Both figures are taken from Ref.~\cite{HALQCD:2012aa}. }
\vskip -0.6cm
\label{fig:potential}
\end{center}
\end{figure}

As an example of results in the HAL QCD method, the leading order NN potential in the isospin-triplet (spin-singlet) channel obtained in 2+1 flavor QCD at $m_\pi \simeq 700$ MeV\cite{HALQCD:2012aa}  is plotted in Fig.~\ref{fig:potential} (left), together with the multi-Gaussian fit,  while the $^1S_0$ scattering phase shift in the laboratory system obtained from this potential  is compared with experimental data in Fig.~\ref{fig:potential}(right).  Both potential and the scattering phase shift well reproduce qualitative features of the nuclear force, though the attraction at low energy is still weaker than experimental one, probably due to the heavier pion mass than physical value $m_\pi \simeq 135$ MeV.

\subsection{Convergence of the derivative expansion}
\begin{figure}[tbh]
\begin{center}
\vskip -0.2cm
  \includegraphics[width=0.5\textwidth]{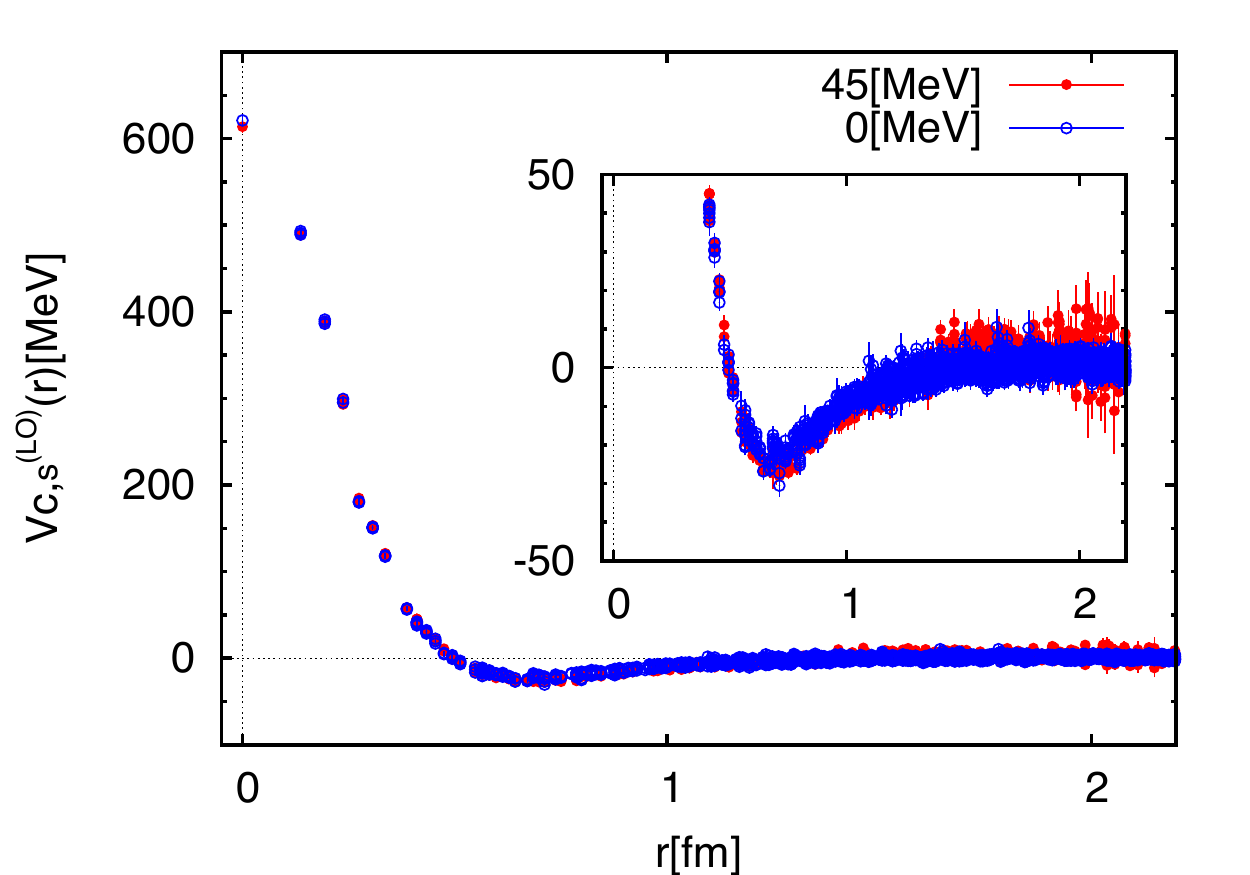}\hfill
  \includegraphics[width=0.5\textwidth]{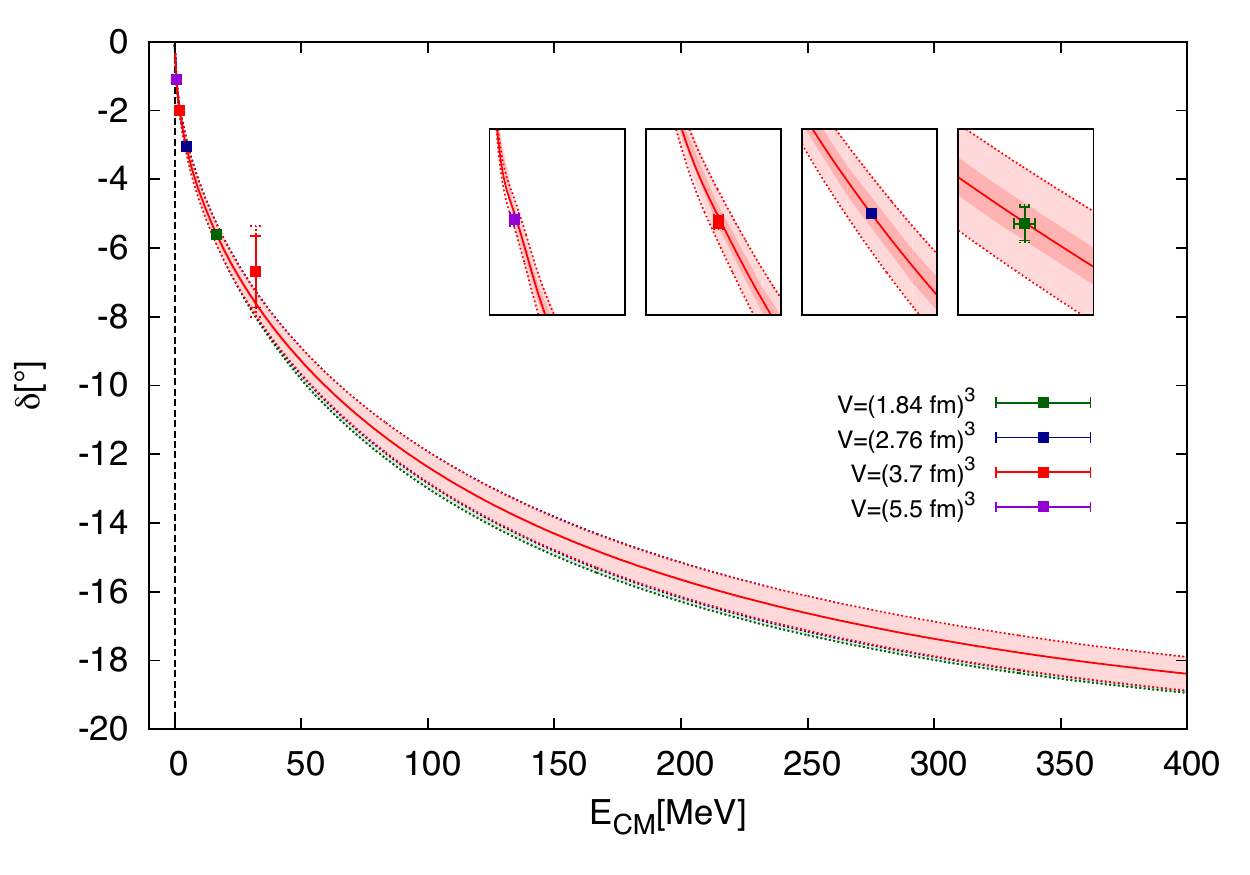}
\caption{(Left) A comparison of the leading order potential in the is spin-singlet channel at  $E_k \simeq 0 $ MeV (blue) and $E_k \simeq 45$ MeV (red) in quenched QCD. A figure is taken from Ref.~\cite{Murano:2011nz}. 
(Right)  Phase shifts of the $I=2$ $\pi\pi$ scattering obtained from the HAL QCD method and L\"usher's finite volume  method in quenched QCD. The red (green) band is obtained by the HAL QCD method at spatial extension $L = 3.7$ (1.8) fm and $E_k\simeq 0$ MeV, while the point data are obtained by the L\"usher's  method. A figure is taken from Ref.~\cite{Kurth:2013tua}.   }
\vskip -1.0cm
\label{fig:convergence}
\end{center}
\end{figure}

If higher order terms in the derivative expansion are large,  the leading order potentials may depend on  energies of the NBS wave functions from which the potential is extracted as in eq.~(\ref{eq:LO}). 
Such truncation ambiguities to the potential have been checked in several cases.
Fig.~\ref{fig:convergence} (left) shows a comparison of the leading order spin-singlet potential  in quenched QCD between two energies in the center of mass system, one is $E_k \simeq 0 $ MeV (blue) and the other is $E_k \simeq 45$ MeV (red)\cite{Murano:2011nz}.  Almost no difference of potentials is seen between two energies, indicating that higher order terms in the derivative expansion turn out to be very small at low energy in the HAL QCD method.
 
Fig.~\ref{fig:convergence} (right)  compares phase shifts of the $I=2$ $\pi\pi$ scattering in quenched QCD calculated from the potential obtained at $E_k\simeq 0$ MeV in the HAL QCD method with those obtained directly at several energies by the L\"usher's finite volume method. As the figure tells, both methods agree extremely well.  This result establishes a validity of the HAL QCD's potential method and shows a good convergence of the derivative expansion.

\section{Discussions}
The HAL QCD method provides an alternative but very powerful method to investigate hadron interactions in (lattice) QCD.  The nuclear potential, calculated in the method, reproduces qualitative features of the nuclear force, not only  medium to long distance attractions but also the short distance repulsion, the repulsive core. 
\begin{figure}[tbh]
\begin{center}
\vskip -0.2cm
  \includegraphics[width=0.5\textwidth]{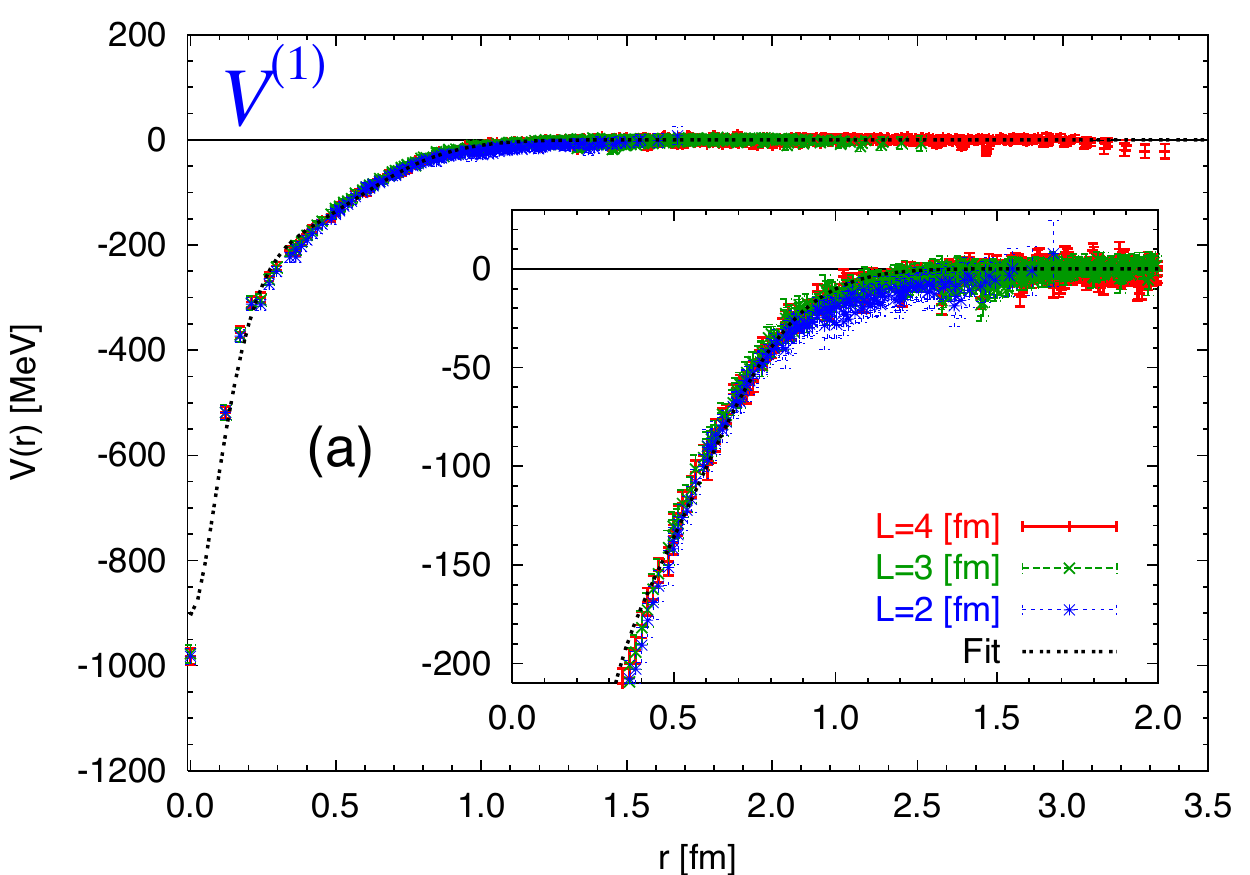}\hfill
  \includegraphics[width=0.5\textwidth]{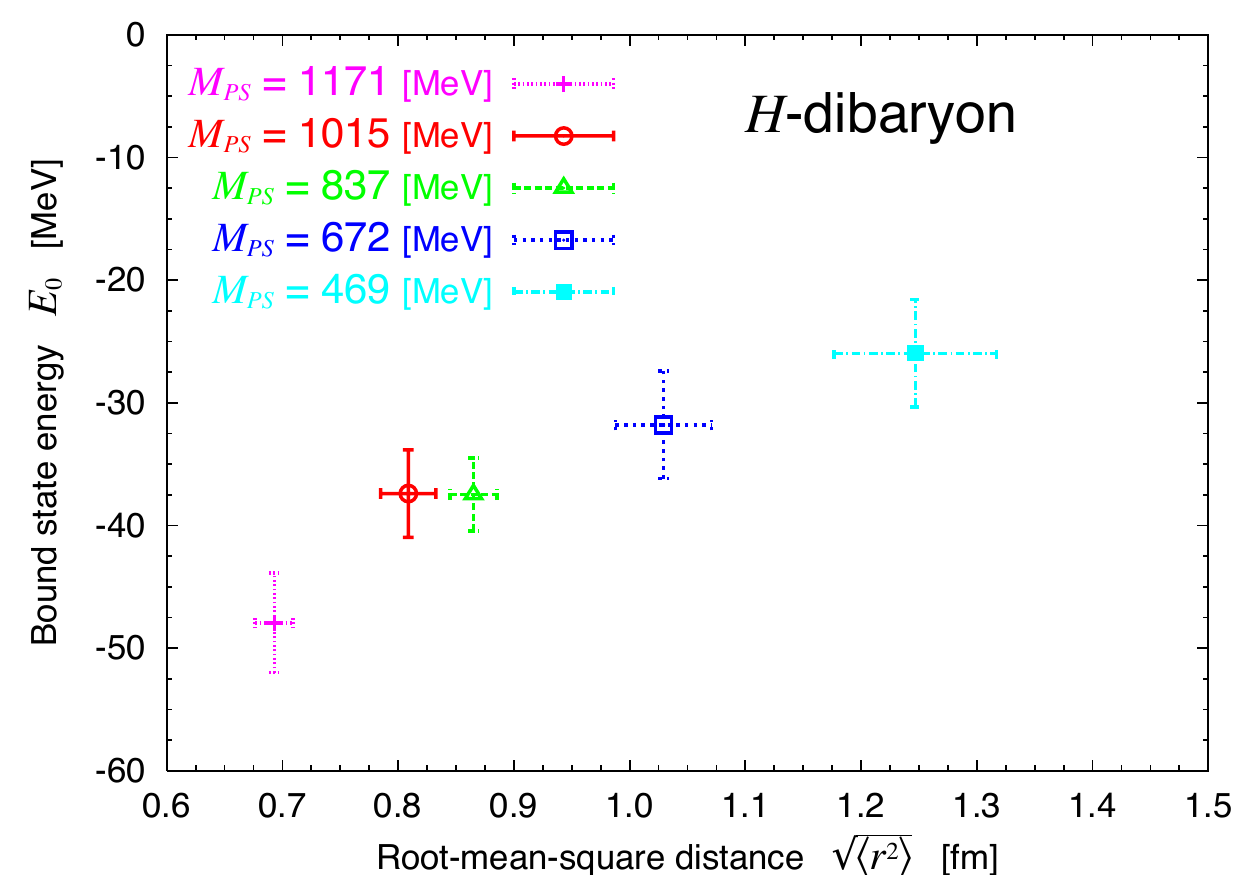}
  \vskip -0.2cm
\caption{(Left) The flavor singlet potential as a function of $r$ on three volumes. A figure is taken from Ref.~\cite{Inoue:2010es}. 
(Right)  The binding energy $E_0$ and the root-mean square distance $\sqrt{\langle r^2\rangle}$ of the H-dibaryon in the flavor SU(3) limit at several values of the pseudo-scalar meson mass. A figure is taken from Ref.~\cite{Inoue:2011ai}.   }
\vskip -0.5cm
\label{fig:singlet}
\end{center}
\end{figure} 

The method can be easily applied to other systems. Fig.~\ref{fig:singlet} (left) shows the SU(3) flavor singlet potential in the 3-flavor QCD with degenerate up-down-strange quarks\cite{Inoue:2010es,Inoue:2011ai}, which has only attractions at all distances. This attraction produces one bond state, H-dibaryon,
whose binding energy  is shown in  Fig.~\ref{fig:singlet} (right). It is interesting to investigate the fate of this state in nature\cite{Sasaki:2012ju}.

\vskip 0.2 cm

This work is supported in part by the
Grant-in-Aid for Scientific Research (25287046), the Grant-in-Aid for Scientific Research on Innovative Areas(No.2004: 20105001, 20105003) and SPIRE (Strategic Program for Innovative Research).

\end{document}